# Intelligent Paging Strategy for Multi-Carrier CDMA System

Sheikh Shanawaz Mostafa[1], Khondker Jahid Reza[2], Md. Ziaul Amin[3] and Mohiuddin Ahmad[4]

[1] Dept. of Biomedical Engineering, Khulna University of Engineering and Technology,
Khulna-9203, Bangladesh.

[2] Electrical & Electronics Engineering, University Malaysia Pahang,
Pahang, Malaysia.

[3] Electronics and Communication Engineering Discipline, Khulna University,
Khulna-9208, Bangladesh

[4] Dept. of Electrical and Electronic Engineering, Khulna University of Engineering and Technology,
Khulna-9203, Bangladesh.

## Abstract

Subscriber satisfaction and maximum radio resource utilization are the pivotal criteria in communication system design. In multi-Carrier CDMA system, different paging algorithms are used for locating user within the shortest possible time and best possible utilization of radio resources. Different paging algorithms underscored different techniques based on the different purposes. However, low servicing time of sequential search and better utilization of radio resources of concurrent search can be utilized simultaneously by swapping of the algorithms. In this paper, intelligent mechanism has been developed for dynamic algorithm assignment basing on time-varying traffic demand, which is predicted by radial basis neural network; and its performance has been analyzed are based on prediction efficiency of different types of data. High prediction efficiency is observed with a good correlation coefficient (0.99) and subsequently better performance is achieved by dynamic paging algorithm assignment. This claim is substantiated by the result of proposed intelligent paging strategy.

## 1. Introduction

In mobile communication, performance of a system depends on the call processing. It is required to identify or page the user to initiate a call for a particular user. The forward-link communication channels which are used to page and transmit system overhead messages to the MS is called paging channel [1].

As personal communication service demands are growing day by day, so it is mandatory to utilize the limited radio resources efficiently. Many researchers have been investigating different mechanisms and techniques to utilize the paging channel efficiently. Among them Rung-Hung Gue and his colleagues [2] proposed concurrent search algorithm. This algorithm was used for paging in cells. Another concept described in [3] and [4] which were related to divide a location area into paging zones. A different type of paging technique has been described in [5] where MS will be paged on the last registered cell first and then other cells in the location area if necessary. A proposal of mobility tracking scheme that merges a movement-based location update policy with a selective paging scheme was proposed in [6]. For efficiently using the resources, probability based priority paging used by other researchers [7]. Neural network has been used for load management in different purpose. A hybrid Bayesian neural network predicting locations on cellular networks was used in [8] and [9]. Neural network for creating personal mobility profile of individual user was used in [10]. Fuzzy logic for paging mobile user was observed in [11]. A Dynamic Location Management for reducing the location update and paging cost was investigated in [12].

In multi carrier CDMA system, an MS is capable of tuning to only one of the carrier frequencies at a time. That's why, it is required that the paging message must be sent on each and every paging channel of all of the carrier frequencies. It is done by duplicating the paging message for each of the paging channels on multicarrier system [13]. As there are only seven paging channels for each carrier in a multi carrier CDMA system, therefore the existing system can't search more than seven users at a time. This result is an inefficient utilization of radio recourses. Concurrent Search Algorithm solves this problem requiring slightly more servicing time compared to sequential search [14].

In this paper, intelligent paging strategy is proposed which can be used in busy hours to search more than seven users simultaneously in multiple carriers, like Concurrent Search Algorithm, and servicing time is as low as Sequential Search Algorithm in normal traffic. We adopt radial basis function (RBF) neural network for predicting paging and choosing the suitable algorithm between two. The system performance has been analyzed by using Erlang C formula.

## 2. Sequential Search Algorithm vs. Concurrent Search Algorithm

This paper is based on the limitation of concurrent search algorithm and Sequential Search Algorithm. In concurrent search algorithm system parallel paging rather than forward sequential paging scheme has been used to locate user i among k number of users.

Presuming there are only two carriers in the system with same number of users. It is also assumed that the system has no previous knowledge about the location of the users: so each user can exist in any of the two carriers with probability 0.5.in sequential paging scheme four pages will be required to locate two users where concurrent search need three messages. Overall 25% saving of paging message means 25% more channel for paging.

If the servicing time is calculated by absorbing Markov chains [15] considering that Search Algorithm has one unit paging (service) time than Concurrent Search Algorithm will have 1.5 units [14]. And if Concurrent system is imagined with 14 parallel servers having queue than Sequential system can be imagined with 7 parallel servers having queue [16]. In this case its average time in the system can be found from Erlang C formula [17], [14] is,

$$T = \frac{A^C}{\mu \times (C-A) \times \left\{ A^C + C!(1-\frac{A}{C})\sum_{k=0}^{C-1}\frac{A^k}{k!} \right\}} + \frac{1}{\mu} \quad (1)$$

And the blocking probability is

$$P_0 = \frac{A^C}{A^C + C!(1-\frac{A}{C})\sum_{k=0}^{C-1}\frac{A^k}{k!}} \quad (2)$$

where $C$ is the number of channels. A is total offered traffic which is $\lambda/\mu$, assuming $\lambda$ is incoming traffic rate and $\mu$ is the service rate.

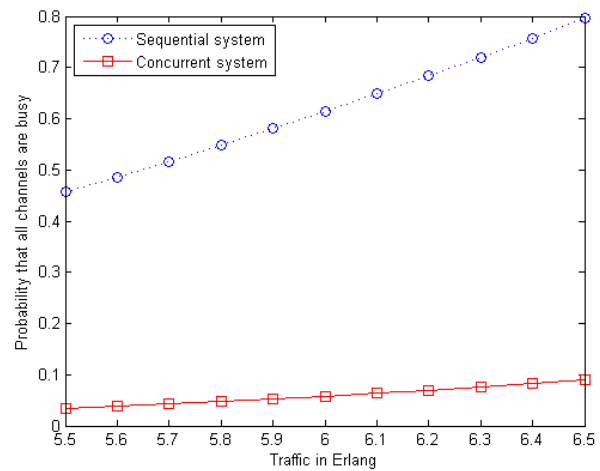

Fig. 1 Blocking probability of Concurrent Search and Sequential Search [14].

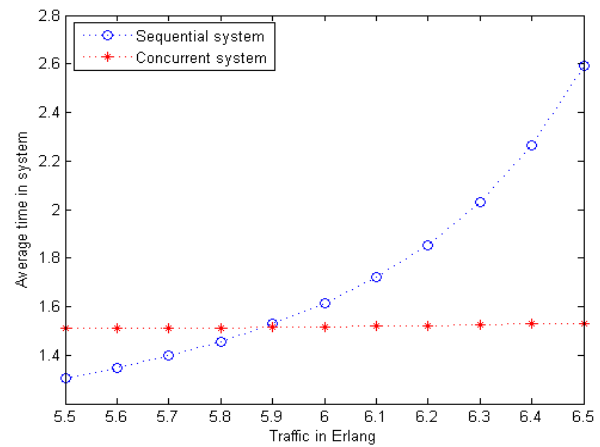

Fig. 2 Servicing time of Concurrent Search and Sequential Search [14].

From Fig. 1 and Fig. 2 it is clear that the blocking probability of concurrent system is always lower than that of sequential system. Considering the system performance in terms of average servicing time, Concurrent Search algorithm will provide low servicing time only after threshold traffic demand, which is six Erlang. Real time traffic analysis may be a possible way out of this problem, but it will introduce computational complexity at the servicing end. Swapping of algorithms based on real time traffic cannot provide feasible solution because the services would be hampered during the switching process of the algorithms. So previously predicted traffic would solve the stated problem. By forecasting the load, the system will know which algorithm is to be used before the arrival of new traffic.

## 3. Method of Intelligent Paging Strategy

### 3.1 Radial Basis Neural Network

The incoming traffic of the mobile system usually follows a specific pattern depending on the characteristics of area and time. A reliable time series prediction method is needed for forecasting this type of system behaviors. In recent years, neural networks gain its recognition among different types of time series prediction methods. Radial basis neural network was used in [18], [19] and [20] for time series prediction problem. Moreover, radial basis function networks have rapid training time and have no local minima problem as back propagation does. Instead of using weighted-sum mechanism or sigmoid activation for hidden-unit outputs, radial basis neural network (RBNN) uses vector distance between its weight vector and the input vector [21]. Radial basis networks consist of two layers of neural (Fig. 3): a radial basis layer (hidden layer), and linear layer.

**Radial Basis Layer:** Each predictor variable defines a neuron in input radial basis layer. In the case of categorical variables, N-1 neurons are used where N is the number of categories. Each neuron's weighted input is the distance between the input vector and its weight vector. Net input is the product of its weighted input with its bias. Each neuron's output h(x) is its net input x passed through the following function.

$$h(x) = e^{\frac{(x-u)^2}{2\sigma^2}} \qquad (3)$$

Where x is the input vector, u is the mean and σ is the standard deviation. [21]

**Linear Layer:** The linear neuron uses a linear transfer function.

$$O = g \times I \qquad (4)$$

Where I is the value coming out of a neuron in the hidden layer is multiplied by a weight associated with the neuron and passed to the summation which adds up the weighted values and presents this sum as the output of the network. For classification problems, there is one output for each target category.

In this paper, three different types of traffic namely type-1 (T1), type-2 (T2) and type-3 (T3) have been considered to simulate the varying characteristics of traffic. The bias in the radial basis layer was set to 0.8326; and neuron will respond with 0.5 or more to any input vectors within a vector distance of one from their weight vector. Mean Square Error (MSE) goal was set to 0.02 to train the radial basis neural network. It was required 50 Neurons to achieve the defined Mean Square Error goal.

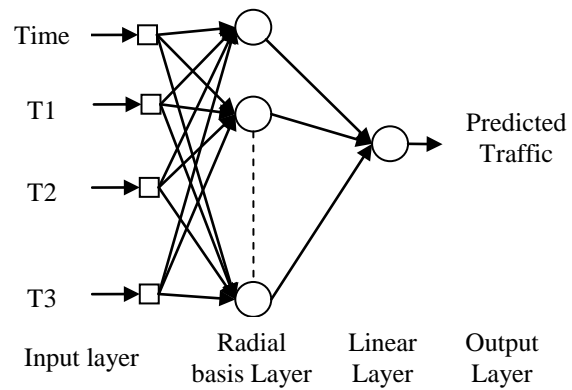

Fig. 3 Radial basis neural network for intelligent paging

### 3.2 Predictor Performance Evaluators

The performance of a predictor can be determined mathematically by calculating the correlation coefficient and error of actual and predicted traffic.

Supposing, x (n) be the actual traffic, and $x_m(n)$ be the predicted traffic generated by the radial basis neural network model, then Mean Square Error (MSE) can be defined by the following equation incorporating them [22].

$$MSE = \frac{1}{N} \sum_{n=0}^{N-1} [x(n) - x_m(n)]^2 \qquad (5)$$

The normalized form of MSE is

$$NMSE = \frac{\sum_{n=0}^{N-1} [x(n) - x_m(n)]^2}{\sum_{n=0}^{N-1} [x(n)]^2} \qquad (6)$$

Another measurement is Root Mean Square Error, which is

$$RMSE = \sqrt{\frac{1}{N}\sum_{n=0}^{N-1}[x(n) - x_m(n)]^2} \qquad (7)$$

The Normalized version of RMSE is

$$NRMSE = \sqrt{\frac{\sum_{n=0}^{N-1}[x(n) - x_m(n)]^2}{\sum_{n=0}^{N-1}[x(n)]^2}} \qquad (8)$$

Percent Root Mean Difference (PRD) can be determined by Eq. (9).

$$PRD = \sqrt{\frac{\sum_{n=0}^{N-1}[x(n) - x_m(n)]^2}{\sum_{n=0}^{N-1}[x(n)]^2}} \times 100\% \qquad (9)$$

Cross Correlation can be defined as a sequence of $r_{xx_m}$,

Where;

$$r_{xx_m}(l) = \sum_{n=-\infty}^{\infty} x(n)x_m(n-l), \quad l=0,\pm1,\pm2,.. \qquad .(10)$$

*or equivalently,*

$$r_{xx_m}(l) = \sum_{n=-\infty}^{\infty} x(n+l)x_m(n), \qquad l=0,\pm1,\pm2,.... \qquad (11)$$

The index $l$ is the time shift parameter and the subscripts $xx_m$ is the correlation sequence and $r_{xx_m}(l)$ the sequences being correlated [23].

## 4. Result and Performance Evaluation

System performance is highly depended on algorithms swapping efficiency, while it depends on the reliability of traffic prediction mechanism. Therefore, the performance of intelligent paging strategy has been analyzed considering the traffic prediction and total system performance which is measured by blocking probability and average servicing time of the system.

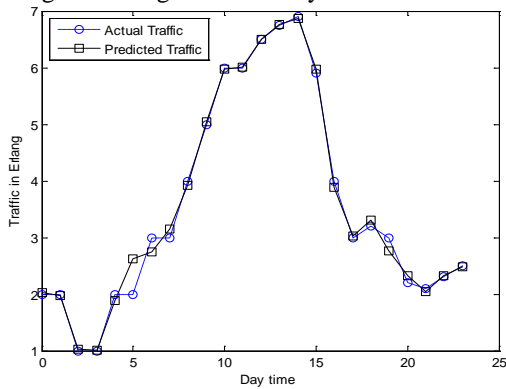

Fig. 4 Prediction of traffic by using RBNN for Type-1 traffic

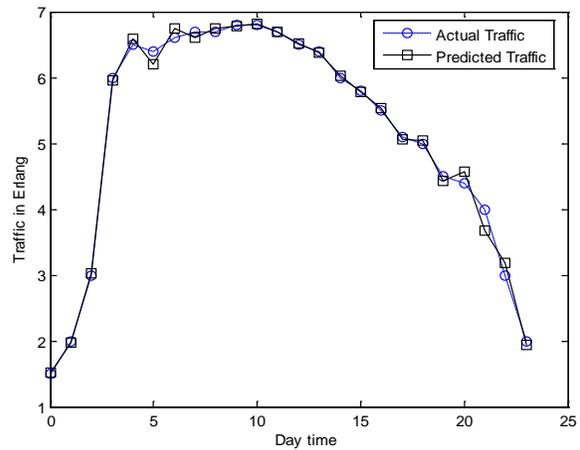

Fig. 5 Prediction of traffic by using RBNN for Type-2 traffic

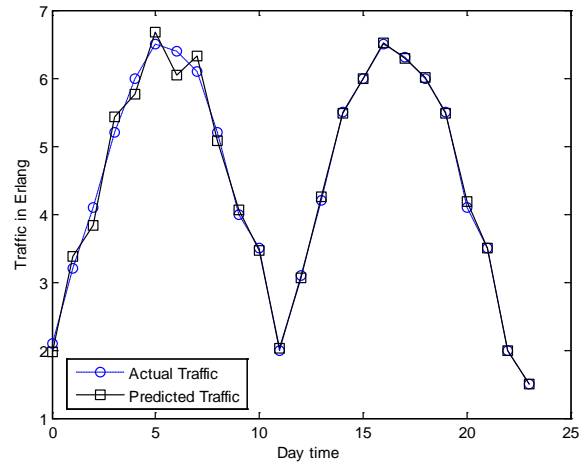

Fig. 6 Prediction of traffic by using RBNN for Type-3 traffic

Since prediction capability will determine the total system performance of the system. So, reliable predictor is required to get a realistic prediction. The Predictor, Neural Network, is trained with random data to get maximum performance. The prediction is almost identical to the simulated traffic with few exceptions and it is substantiated by the Fig 4, Fig 5 and Fig 6.

Mathematical error performance of predictor is presented in Table I. It is clearly depicted the performance of predictor by the high Cross correlation value and low MSE, RMSE and PRD.

Table I: Performance parameters of the predictor

|    | MSE | NMSE | RMSE | NRMSE | PRD | Correlation Coefficient |
|----|-----|------|------|-------|-----|-------------------------|
| T1 | 2.6e-2 | 1.6e-3 | 1.6e-1 | 2.5e-6 | 2.5e-4 | 0.9992 |
| T2 | 1.1e-2 | 3.7e-4 | 1e-1 | 1.4e-7 | 1.3e-5 | 0.9998 |
| T3 | 2e-2 | 8.8e-4 | 1.4e-1 | 7.8e-7 | 7.8e-5 | 0.9996 |

System Performance depends on timely swapping of algorithm. This is achieved due to good prediction of the predictor and successful swapping is seen in the Fig 7, Fig 8 and Fig 9.

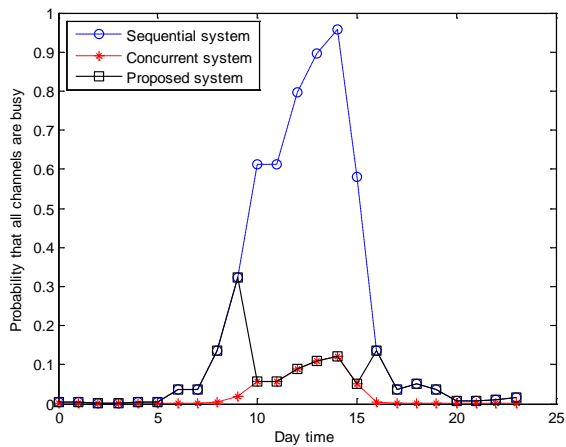

Fig. 7 Blocking probability comparison of proposed, concurrent and sequential search for Type-1 traffic.

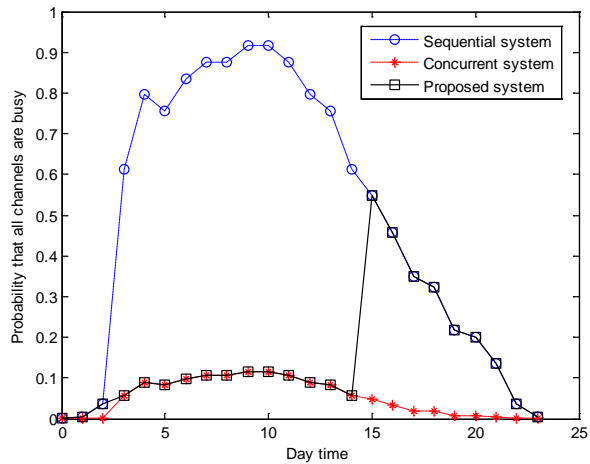

Fig. 8 Blocking probability comparison of proposed, concurrent and sequential search for Type-2 traffic

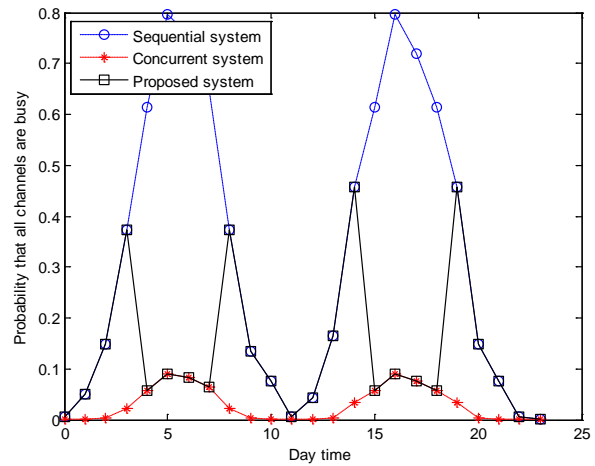

Fig. 9 Blocking probability comparison of proposed, concurrent and sequential search for Type-3 traffic

Performance of the system, which is least possible service time, can be achieved by efficient swapping of the two algorithms. When traffic is lower than the threshold traffic, sequential search will require minimal time. That is why our intelligent system will follow sequential search algorithm. When traffic is greater than the threshold traffic, then the minimal accessing time is provided by concurrent search algorithm. From that time our intelligent paging switches to concurrent search algorithm. The reverse pattern is also observed in the Fig 10, Fig11and Fig 12.

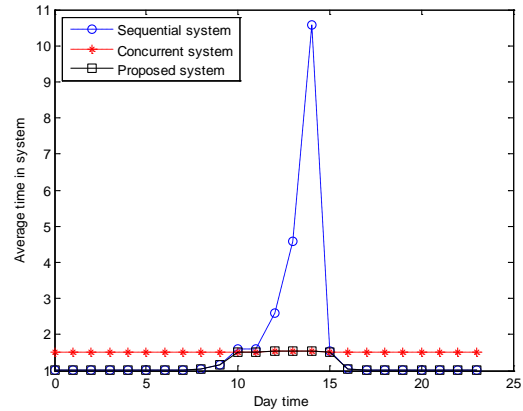

Fig. 10 Servicing time comparison of proposed, concurrent and sequential search for Type-1 traffic.

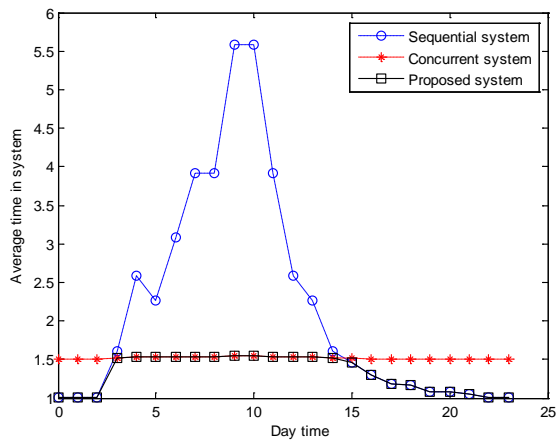

Fig. 12  Servicing time comparison of proposed, concurrent and
sequential search for Type-3 traffic.

## 5. Conclusion

In this paper, an intelligent paging strategy has been developed assuming that traffic follows some deterministic pattern which is predicted by radial basis neural network. It not only forecast the traffic pattern with less than 0.15 RMSE error but also increases the system performance in varying traffic demand which is rather common in wireless system. Due to offline training method implemented in this strategy, the training time of the network has no significant effect on the system. It requires very limited resources of the implementing terminal. However, this strategy has been applied on simulated traffic with good result. The model can be further studied comprehensively by using real traffic data.

**Sheikh Shanawaz Mostafa** completed B.Sc. Engineering in Electronics and Communication, from Khulna University-9208, Khulna, Bangladesh. His current research interests are: Wireless communication, Modulation techniques and Biomedical signal processing. He has more than seven papers, published in different local and international recognized journal and proceedings of conference.

**Khondker Jahid Reza** has completed his B.Sc. in Electronics and Communication Engineering Discipline in Khulna University, Khulna, Bangladesh. His current research interest is wireless communication, modulation and sensor networks. He has three papers, published in international recognized journal.

**Md. Ziaul Amin** is currently working as an Assistant Professor at the Khulna University, Khulna, Bangladesh. He obtained his B.Sc. in Electronics and Communication Engineering from same University. Previously, he worked as a System Engineer, planning, at RanksTel Bangladesh Ltd. since Nov 07 Aug 08. His current research interests are: Digital Signal Processing, Radio Network Planning and cognitive radio network. He has four papers, published in international recognized journal.

**Dr. Mohiuddin Ahmad** received his BS degree with Honors in Electrical and Electronic Engineering from Chittagong University of Engineering and Technology (CUET), Bangladesh and his MS degree in Electronics and Information Science from Kyoto Institute of Technology of Japan in 1994 and 2001, respectively. He received his PhD degree in Computer Science and Engineering from Korea University, Republic of Korea. From August 1995 to October 1998, he served as a lecturer in the Department of Electrical and Electronic Engineering at Khulna University of Engineering and Technology, Bangladesh. In June 2001, he joined the same Department as an Assistant Professor. In May 2009, he joined the same Department as an Associate Professor and now he is a full Professor. His research interests include biomedical signal and image processing, computer vision and pattern recognition, human motion analysis, circuits and energy conversion.